\documentstyle[12pt]{article}
\begin{document}
\begin{titlepage}
\pagestyle{empty}
\baselineskip=21pt
\rightline{Alberta Thy-31-92}
\rightline{LBL-32888}
\rightline{UMN-TH-1113-92}
\rightline{September 1992}
\vskip .2in
\begin{center}
{\large{\bf Supersymmetric (S)Neutrino-Mass Induced Baryogenesis}}
\end{center}
\vskip .1in
\begin{center}
Bruce A. Campbell

{\it Department of Physics, University of Alberta}

{\it  Edmonton, Alberta, Canada T6G 2J1}

Sacha Davidson

{\it Center for Particle Astrophysics, University of California}

{\it Berkeley, California, 94720 USA}

Keith A. Olive

{\it Department of Physics and Astronomy, University of Minnesota}

{\it Minneapolis, MN 55455, USA}

\vskip .2in

\end{center}
\centerline{ {\bf Abstract} }
\baselineskip=18pt

\noindent

We propose a new mechanism for baryogenesis in supersymmetric
extensions of the standard model, that does not depend on (super)GUT
interactions. It occurs by the non-perturbative electroweak
reprocessing of a lepton asymmetry. This lepton asymmetry is
generated by the effects of lepton number violating induced
operators, arising from ``see-saw" (s)neutrino masses, which act on
scalar condensate oscillations along flat directions of the
supersymmetric standard model.
\vspace*{1cm}

\centerline{ {Accepted For Publication In: {\it Phys.Lett. B}} }

\vspace*{1cm}

\end{titlepage}
\baselineskip=18pt
{\newcommand{\la}{\mbox{\raisebox{-.6ex}{$\stackrel{<}{\sim}$}}}
{\newcommand{\ga}{\mbox{\raisebox{-.6ex}{$\stackrel{>}{\sim}$}}}

       The baryon asymmetry of the universe represents one of the
basic facts of cosmology, which must be addressed by any theory of
the fundamental interations.
This is most emphatically the case in the context of inflationary
models of the universe \cite{infl}, where any preexisting baryon
asymmetry would have been inflated away, and we are forced to
rely on the microphysics of elementary particle interactions to
generate the observed BAU. The conditions necessary for this to occur
as enunciated by Sakharov \cite{sak} are: non-conservation of baryon
number; violation of C and CP symmetry, and out of equilibrium
dynamics.

The first realizations of the Sakharov conditions were in the
context of Grand Unified Theories (GUTs). In these theories, quarks
and leptons were unified into multiplets of a larger simple gauge
symmetry, which contained the Standard Model gauge group as a
subgroup. Gauge and Higgs interactions of the larger simple group
then
violated baryon (B) and lepton (L) number at the high energy scales
where grand unification occurred. During the thermal history of the
Universe, heavy gauge and Higgs particles could, via their
out-of-equilibrium decays, generate cosmological baryon and lepton
asymmetries \cite{kt}.The constraints on such a scenario derive from
two conflicting requirements. To generate a BAU after inflation
requires that the relevant dynamics occur at energy scales below
either the inflaton mass scale, or the reheating temperature. On the
other hand, GUT reactions of the type responsible for generation of
the BAU are generally the same interactions that cause proton decay;
the
non-observation of the latter puts lower limits on the energy scale
at which this dynamics can occur. In many models, proton stability
considerations are difficult to reconcile with baryogenesis after
inflation.

       New possibilities arose after the realization \cite{krs1} that
at temperatures above the electroweak transition temperature baryon
number violating nonperturbative electroweak interactions
(sphalerons) would be in equilibrium. Since these interactions
conserve B-L, it would suffice to produce a net lepton asymmetry, and
then rely on the sphaleron effects to (partially) convert it to a
BAU. The great advantage here is that one no longer needs GUT
dynamics, and the concomitant problems of proton stability. Fukugita
and Yanagida proposed a scenario \cite{fy1} where the lepton
asymmetry is generated in the out of equilibrium decay of a heavy
right handed neutrino $\nu_R$, as arises in the see-saw \cite{seesaw}
mass matrix
that gauge symmetries  permit for models with a $\nu_R$. The
Fukugita-Yanagida mechanism is similar to the standard GUT scenario
in that it utilizes out of equilibrium decay as its dynamical
mechanism; it is distinct and original in that it utilizes neutrino
Majorana masses, not GUT interactions, as its source of B-L
violation, producing only a lepton asymmetry for later sphaleron
reprocessing.

       In supersymmetric Grand Unified Theories (susyGUTs) the baryon
number violating GUT interactions can also effect a BAU by their
effects on scalar superpartner fields \cite{ad,lin85,ceno,eeno,morg}.
As Affleck and Dine \cite{ad} first proposed, GUT interactions induce
an effective contribution to the scalar potential (after
supersymmetry breaking) which could act on the oscillations of
coherent sfermion condensates so as to generate a BAU. The dynamics
are entirely different, but the underlying interaction is still that
of GUT violation of baryon number.

       In this paper we propose a new mechanism of baryogenesis,
arising from the presence of see-saw neutrino masses in
supersymmetric theories; no GUT interactions are required. We first
derive the contributions induced after supersymmetry breaking, in the
low energy effective potential of a supersymmetric theory, from the
effects of see-saw neutrino mass generation. We then examine the
effect of these interactions on slepton and squark condensates
oscillating along ``flat directions" (before supersymmetry breaking)
of the low-energy supersymmetric standard model. We show that these
interactions act on the condensate oscillations to produce a net
lepton asymmetry. Sphaleron reprocessing of the lepton asymmetry to
baryons then completes our scenario. Our mechanism resembles that of
Fukugita and Yanagida in that the only ``non-standard" interaction
required is the see-saw neutrino mass, and in that the dynamical
mechanism only generates a lepton asymmetry at first, with sphalerons
responsible for partially reprocessing that into baryons.

       On the other hand, unlike the proposal of Fukugita and
Yanagida, the dynamics of our mechanism involves the oscillations of
sfermion condensates along susy flat directions, and hence
unlike theirs can only occur in supersymmetric extensions of the
standard model. Furthermore, in our mechanism the CP violation
necessary for the production of the asymmetry may arise spontaneously
from the phase of the condensate, and hence is naturally of order
unity, whereas in the Fukugita-Yanagida scenario one needs hard CP
violation in the neutrino-Higgs Yukawa couplings. Finally, as our
mechanism depends on the effective low energy interactions induced by
the right handed see-saw neutrinos, it can be operative even when the
$\nu_R$ masses are too large for them to be physically produced in
the
post inflationary epoch, whereas in the Fukugita-Yanagida mechanism
they must be copiously produced, thus bounding their mass by the
inflaton mass scale, which COBE results give as $\leq 10^{11} GeV$
for typical inflationary models \cite{cdo2}.

       In order to demonstrate our mechanism of baryogenesis we need
the following elements. First we need to show the existence of flat
(before supersymmetry breaking) directions in the potential of our
model, including the contributions from the F-terms involving the singlet
$\nu^c$. It is the scalar condensates
(squark, slepton, and Higgs) along these directions whose motion,
after
supersymmetry breaking potentials turn on, drives the lepton
asymmetry generation. Second, we need to establish the existence of
slepton number violating potential interactions, induced after
supersymmetry breaking by the singlet neutrino superpotential
terms, which pick up a non-zero contribution along the flat
direction, and which act during the course of the scalar oscillations
to build up a net slepton density. Third, we must follow the
evolution of the condensate to calculate the lepton asymmetry
produced, and its subsequent dilution by inflaton decay, to get the
final lepton asymmetry for sphaleron reprocessing.

       To begin our demonstration example, we need to exhibit flat
directions in the supersymmetric standard model, extended to include
neutrino see-saw masses and Higgs mixing.
 The superpotential terms contributing to the
potential are:
\begin{eqnarray}
W & = & {g_u^{ij}}({H_1}{Q^i}{u^{cj}})+{g_d^{ij}}({H_2}{Q^i}{d^{cj}})
+{g_e^{ij}}({H_2}{L^i}{e^{cj}}) + m_H H_1 H_2 \\ \nonumber
&&+{g_{\nu}^{ij}}({H_1}{L^i}{{\nu}^{cj}}
) +{M^{ij}}{{\nu}^{ci}}{{\nu}^{cj}} + k_1^i(\nu^{ci}H_1H_2) + k_2^{ijk}
(\nu^{ci}\nu^{cj}\nu^{ck})
\end{eqnarray}
where the left chiral supermultiplets are labelled by their
particles, and $H_1$ and $H_2$ are the two Higgs doublet
supermultiplets.
 The potential resulting from these F-terms (plus the
SU(3)xSU(2)xU(1) D-terms) has the following useful flat direction
(which is a generation permuted version of one appearing in reference
\cite{morg}, and which may be shown to acquire no new contributions
from F-terms associated with the neutrino mass see-saw); it depends
on three arbitrary complex parameters $a,v,c,$ and four phases
$\alpha,\beta,
\phi$, and $\gamma$. We work in a generation basis in which the
${g_e^{ij}}$ and the ${g_d^{ij}}$ have been diagonalized; the quark
indices denote quark colour.
\begin{equation}
\begin{array}{ccc}
{{\tilde{t}}^c_3}=a & {{\tilde{t}}^1}=v &
{{\tilde{\nu}}_e}={e^{i\gamma}}c\\
{{\tilde{b}}^c_3}=c &{{\tilde{s}}^c_2}={e^{i\alpha}}
\sqrt{{|a|}^2+{|c|}^2}& \\
{{\tilde{\mu}}^-}={e^{i\beta}} \sqrt{{|v|}^2+{|c|}^2}
&{{\tilde{d}}^c_1}={e^{i\phi}} \sqrt{{|a|}^2+{|v|}^2+{|c|}^2}& \\
\end{array}
\end{equation}
This particular flat direction has the virtue that its vev produces a
non-zero value for the effective scalar operator
\begin{equation}
\langle{{\tilde{\mu}}^-}{{\tilde{\nu}}_e}{{\tilde{b}}^c_3}{({{\tilde{
t}}^c_3})^*}\rangle = {e^{i(\beta+\gamma)}}{a^*}{c^2}
\sqrt{{|v|}^2+{|c|}^2}
\label{a}
\end{equation}
which violates lepton number by two units.

       After supersymmetry breaking, this scalar operator will be
induced by the neutrino see-saw couplings via the diagram of Figure
1. In the diagram the insertions on the $\tilde{\nu}^c$ line and vertex
are the
supersymmetry breaking scalar mass and interaction A-terms ($O(m_{\delta}$)).
 The
resulting potential term coupling is of order
$V=\lambda{{\phi}{\phi}{\phi}{{\phi}^*}}$,
where ${{\phi}{\phi}{\phi}{{\phi}^*}}$ corresponds to the quartic
scalar operator of equation (\ref{a}) and
\begin{equation}
\lambda \sim \frac{{g_{\nu}^e}{g_{\nu}^{\mu}}{g_b}{g_t}}{(4\pi)^3}
\frac{m_{\delta}^2 M ^2}{(M^2 + g^2\phi_o^2)^2}
\label{b}
\end{equation}
where ${g_{\nu}^e}$ and ${g_{\nu}^{\mu}}$ are the (experimentally
undetermined) neutrino
see-saw Dirac mass Yukawas, and M is the scale of the large singlet
${\nu}^c$ Majorana mass term. (We assume that $k_1^i$and  $k_2^{ijk}$
are of order 1).  We allow the possibility that
the scalars $\phi$ pick up expectation values larger than M, giving effective
loop propagator masses $O(g \phi_o)$ on legs for which this exceeds the direct
mass.
 The estimate for the scale of the
induced quartic scalar coupling agrees with the general arguments of
\cite{morg} for operators of the form ${\phi}{\phi}{\phi}{{\phi}^*}$.

       So we have: a flat direction; a see-saw generated lepton
number violating quartic scalar potential interaction, which is
induced along the flat direction after supersymmetry breaking; and
an arbitrary phase for the flat direction vevs which breaks CP
spontaneously and is generically of order unity. It remains only
to calculate the net lepton (and baryon) asymmetry produced by the
the decay of the sfermion oscillations.

       After inflation, denote the expectation values of scalars
parametrizing the flat directions as $\phi_o
=\langle{o|\phi|o}\rangle$, producing $V_o = \langle{o|V|o}\rangle$.
We can then write the net lepton number per scalar particle
associated
with the oscillations of $\phi$ as
\begin{equation}
L \sim \frac{Im V_o}{{m_\delta}^2 {\phi_o}^2}
\label{c}
\end{equation}
 From equations (\ref{a}) and (\ref{b}) we see that
\begin{equation}
Im V_o \sim \theta \lambda {\phi_o}^4
\label{d}
\end{equation}
where $\theta \sim 1$ is the degree of CP violation in (\ref{a}).
Thus we can
write the net lepton number per particle as
\begin{equation}
L \sim O(10^{-5}) \theta g_{\nu}^e g_{\nu}^\mu \frac{ M ^2
{\phi_o}^2}{(g^2\phi_o^2 + M^2)^2}
\label{e}
\end{equation}
where we have assumed that $g_t \sim 1$ and
$g_b \sim O(10^{-2})$.
The net lepton number density ($\sim$ the net baryon density after
sphaleron
reprocessing) is then given by
\begin{equation}
n_B \sim n_L \sim L m_\delta {\phi_o}^2 {(R_\phi/R)}^3
\label{f}
\end{equation}
where $R$ is the cosmological scale factor and $R_\phi$ is the value
of the scale factor when the sfermion oscillations begin.

In the absence of inflation, the final baryon asymmetry is easily
found using \cite{ad}
\begin{equation}
\frac{n_B}{n_\gamma} \sim O(10^{-5}) \theta
\frac{ M ^2
{\phi_o}^2}{(g^2\phi_o^2 + M^2)^2}
{( g_{\nu}^e g_{\nu}^\mu\frac{M_P}{m_\delta})}^{1/6}
\end{equation}
which as one can see may be quite large (though this is known to be an
overestimate \cite{lin85}).
In the context of inflation, it is known that the final asymmetry is
generally much
smaller \cite{eeno,cdo2}.
First, the initial value of $\phi_o$ is determined by quantum
fluctuations during
inflation, and second there is the dilution of the asymmetry due to
inflaton decays.
If we assume a single mass scale $\mu$ to be associated with
inflation (as can be
done for most simple inflationary models), then recent COBE
observations of the
microwave background anisotropies \cite{cobe} indicate that $\mu^2 =
few \times 10^{-8}
{M_P}^2$ \cite{cdo2}. The initial value for the sfermion expectation
value
is ${\phi_o}^2 = H^3\tau/4\pi$ where the Hubble parameter  $H \simeq
\mu^2/M_P$ and
the duration of inflation is $\tau \simeq {M_P}^3/\mu^4$ so that
$\phi_o \simeq \mu$ \cite{cdo2}.
The final baryon asymmetry can then be found from \cite{eeno} to be
\begin{equation}
\frac{n_B}{n_\gamma} \simeq O(10^{-5}) \frac{\theta g_{\nu}^e
g_{\nu}^\mu M^2 {\phi_o}^4
 {m_\eta}^{3/2}}{(M^2 + g^2\phi_o^2)^2 {M_P}^{5/2}
 m_\delta} \simeq  O(10^{-2}) \frac{
\theta g_{\nu}^e g_{\nu}^\mu \mu^3 M^2}{{M_P}^4 m_\delta}
\label{g}
\end{equation}
where $m_\eta \simeq \mu^2/M_P$ is the inflaton mass. We have
assumed $M \la g\phi_o$ and $g^4 \sim g_t^3g_b \sim 10^{-3}$ in the
denominator.  Equation
(\ref{g}) will yield a
required value of $\simeq 10^{-10}$ for $\theta g_{\nu}^e g_{\nu}^\mu
M^2/M_P^2 \simeq {10^{-13}}$.

We can now see clearly the difference between this scenario and
previous scenarios for
generating the baryon asymmetry. First as we stressed earlier this
scenario does not
depend in any way upon grand unification as does the original
mechanism proposed by
Affleck and Dine \cite{ad}.
It works in the absense of any GUT baryon or
lepton number violation.  It also differs significantly from the
heavy lepton decay
scenario of Fukugita and Yanagida \cite{fy1}.  In their scenario, the
baryon asymetry
is a reprocessed lepton asymmetry due to the decay of the heavy
right-handed neutrino
with mass $\simeq M$. The mass scale M must be small enough so that
the right handed
neutrino is produced in inflaton decays $M \leq m_\eta \simeq 10^{11}
GeV$\cite{cdo2}, yet not too
small so that the lepton asymmetry is actually erased due to induced
dimension-five
operators\cite{fy2}.
In our case, there is little restriction beyond that due to the
dimension-five
operator which in this case is $M \geq g_{\nu}^2 10^{11} GeV$ \cite{cdo2}.
Indeed our scenario works better for large values of M and can
thus be thought of as complementary to the Fukugita-Yanagida as
well as Affleck-Dine mechanisms.

       In summary, we have proposed a new mechanism for baryogenesis
in supersymmetric extensions of the standard model, that does not
depend on (super)GUT interactions. It depends on non-perturbative
electroweak reprocessing of a lepton asymmetry. The lepton asymmetry
is generated by the effects of lepton number violating induced
operators, acting on scalar condensate oscillations along flat
directions of the standard model. In our realization of the mechanism
we have shown that lepton number violating operators of this type can
be induced, after supersymmetry breaking, by  singlet neutrino interactions
that include a Majorana mass
term. But the idea is quite a general feature of
supersymmetric extensions of the standard model. Any source of
violation of either baryon or lepton number has, in principle, the
potential to induce baryogenesis via sfermion condensate dynamics,
coupled with sphaleron reprocessing, in a manner similar to that of
the example we have presented. As such, we expect that in
supersymmetric models, whatever the sources of lepton or baryon
number violation, GUT or otherwise, their effects on flat direction
sfermion oscillations may quite generally yield a possible mechanism
for generating the lepton and baryon asymmetries of the universe.

\vskip .1in

\noindent{ {\bf Acknowledgements} } \\
\noindent  The work of BAC and SD was supported in part by the
Natural Sciences and Engineering Research Council of Canada. The work
of KAO was supported in part by DOE grant DE-AC02-83ER-40105, and by
a Presidential Young Investigator Award.
}
}

\newpage

\noindent{\bf{Figure Captions}}

\vskip .1in

\noindent Figure 1:  Diagram inducing lepton number violation in the
low energy effective potential, after supersymmetry breaking.


\newpage

\begin{thebibliography}{99}
\bibitem{infl}for reviews see: A.D. Linde, $\underline{Particle~
Physics~And~Inflationary~Cosmology}$, Harwood (1990); K.A. Olive,
Phys. Rep.
{\bf C190}(1990)307.
\bibitem{sak}A.D. Sakharov, JETP Lett.{\bf 5}(1967)24.
\bibitem{kt}for reviews see: E.W. Kolb and M. Turner, Ann. Rev. Nucl.
Part. Sci.{\bf 33}(1983)645; A. Dolgov, Yukawa Institute preprint
YITP/K-940, 1991.
\bibitem{krs1}V. Kuzmin, V. Rubakov, and M.Shaposhnikov, Phys. Lett.
{\bf B155}(1985)36.
\bibitem{fy1}M. Fukugita and T. Yanagida, Phys. Lett.{\bf
B174}(1986)45.
\bibitem{seesaw}M. Gell-Mann, P. Ramond, and R. Slansky, in
$\underline{Supergravity}$, eds. D.Z. Freedman and P. van
Nieuwenhuizen,
North Holland (1979); T. Yanagida, in Proceedings of the Workshop on
the
Unified Theory and The Baryon Number of the Universe, eds O. Sawada
and
S. Sugamoto. KEK79-18 (1979).
\bibitem{ad}I. Affleck and M. Dine, Nucl. Phys. {\bf B249}(1985)361.
\bibitem{lin85}A.D. Linde, Phys. Lett, {\bf B160}(1985)243.
\bibitem{ceno}B.A. Campbell, J. Ellis, D.V. Nanopoulos, and K.A.
Olive, Mod. Phys. Lett. {\bf A1}(1986)389.
\bibitem{eeno}J. Ellis, K. Enqvist, D.V. Nanopoulos, and K.A. Olive,
Phys. Lett. {\bf B191}(1987) 343.
\bibitem{morg}D. Morgan, Nucl. Phys. {\bf B364}(1991)401.
\bibitem{cdo2}B.A. Campbell, S. Davidson, and K.A. Olive,  Alberta Thy-38-92,
CfPA 92-035, UMN-TH-1114/92.
\bibitem{cobe}F.L. Wright et al. Ap. J. Lett. {\bf 396}(1992)L13.
\bibitem{fy2}M. Fukugita and T. Yanagida, Phys. Rev.{\bf
D42}(1990)1285.
\end{thebibliography}
\end{document}